%% file: sn-article.tex
\documentclass[pdflatex,sn-mathphys-num]{sn-jnl}

\usepackage{graphicx}%
\usepackage{multirow}%
\usepackage{amsmath,amssymb,amsfonts}%
\usepackage{amsthm}%
\usepackage{mathrsfs}%
\usepackage[title]{appendix}%
\usepackage{xcolor}%
\usepackage{textcomp}%
\usepackage{manyfoot}%
\usepackage{booktabs}%
\usepackage{bookmark}
\usepackage{algorithm}%
\usepackage{algorithmicx}%
\usepackage{algpseudocode}%
\usepackage{listings}%

\usepackage[acronym]{glossaries}
\usepackage{anyfontsize}

\theoremstyle{thmstyleone}%
\theoremstyle{thmstyletwo}%

\theoremstyle{thmstylethree}%

\begin{document}
\input{acronym}
\title[A Gated Hybrid Collaborative Filtering Recommendation]{A Gated Hybrid Contrastive Collaborative Filtering Recommendation}


\author*[1]{\fnm{Eduardo} \sur{Ferreira da Silva}}\email{eduardoferreira@ufba.br}
\equalcont{These authors contributed equally to this work.}

\author[1]{\fnm{Mayki} \sur{dos Santos Oliveira}}\email{maykioliveira@ufba.br}
\equalcont{These authors contributed equally to this work.}

\author[1]{\fnm{Joel} \sur{Machado Pires}}\email{joelpires@ufba.br}
\equalcont{These authors contributed equally to this work.}

\author[1]{\fnm{Denis} \sur{Dantas Boaventura}}\email{denis.boaventura@ufba.br}
\equalcont{These authors contributed equally to this work.}

\author[1]{\fnm{Maycon} \sur{Leone Maciel Peixoto}}\email{maycon.leone@ufba.br}
\author[1]{\fnm{Cassio} \sur{Vinicius Serafim Prazeres}}\email{prazeres@ufba.br}

\author[1]{\fnm{Gustavo} \sur{Bittencourt Figueiredo}\email{gustavobf@ufba.br}}
\author[2]{\fnm{Miriam} \sur{Capretz}}\email{mcapretz@uwo.ca}

\author[1]{\fnm{Frederico} \sur{Araujo Durão}}\email{fdurao@ufba.br}
\equalcont{These authors contributed equally to this work.}

\affil*[1]{\orgdiv{Institute of Computing}, \orgname{Federal University of Bahia}, \orgaddress{ \city{Salvador},  \state{Bahia}, \country{Brazil}}}

\affil[2]{\orgdiv{Department of Electrical and Computer Engineering}, \orgname{Western University}, \orgaddress{\city{London}, \state{Ontario}, \country{Canada}}}


\abstract{Recommender systems increasingly incorporate textual reviews to enrich user and item representations. However, most review-aware models remain optimized for rating prediction rather than ranking quality. This misalignment limits their effectiveness in top-N recommendation scenarios, where discriminative ranking is essential. To address this gap, we propose a Gated Hybrid Collaborative Filtering framework that integrates review-derived representations into an autoencoder-based collaborative model. The architecture injects semantic signals layer-wise through an adaptive gating mechanism that dynamically balances collaborative embeddings and topic-based features during encoding.
To further refine the latent space, we introduce a contrastive learning module that aligns semantic and collaborative signals. We evaluate the framework across five distinct configurations: Pure collaborative; Topic and Gated; Text and Gated; and the addition of contrastive objectives (Contrastive and Topic, and Contrastive and Text).
To explicitly optimize ranking behavior, the model is trained with a pairwise Bayesian personalized ranking objective, which promotes separation between relevant and non-relevant items in the latent space. Experiments on Amazon Movies \& TV, IMDb, and Rotten Tomatoes demonstrate consistent improvements in hit rate @10 and normalized discounted cumulative gain @10 over state-of-the-art review-aware baselines. Results highlight the importance of controlled semantic fusion for ranking-driven recommendation.}

\maketitle

\section{Introduction}
\label{introduction}

Recommender systems traditionally rely on structured interaction signals, such as ratings, clicks, and user–item histories, which often compress complex user experiences into limited representations, such as a single numerical score. To overcome this limitation, recent research has increasingly incorporated textual reviews, which provide richer, multidimensional insights into user preferences that are not captured by structured interaction data~\cite{Tao_Chen_2022, Zheng_Li_2023, Qiang_Wang_2022, Xi_Wang_2021}. Beyond improving representation quality, reviews help mitigate sparsity and cold-start issues by offering additional semantic evidence about users and items, and they support explainability by grounding recommendations in user-generated feedback~\cite{Adomavicius_2005, Khan_2023}.

Advances in natural language processing further enable the extraction of aspects, sentiments, and latent topics, enriching user and item profiles in recommendation models~\cite{Li_Chen_2015, Gheewala_2024, Zhuang_2021}.
However, the expressiveness of reviews introduces important modeling challenges. Given their unstructured, subjective, and often noisy nature~\cite{Raza_2022}, reviews require robust strategies for representation and integration. Beyond influencing individual decisions, they also shape collective perceptions and market dynamics, while aggregate indicators such as average ratings affect consumer judgment and business feedback processes~\cite{Tao_Chen_2022}. Although diverse neural architectures, including convolutional and attention-based mechanisms, have been proposed to incorporate textual information, a structural limitation persists in review-based recommendation research.

Specifically, most review-aware models continue to optimize rating prediction accuracy (e.g., \gls{rmse}, \gls{mae}) rather than ranking quality, implicitly assuming that improved numerical estimates lead to better recommendations~\cite{Zheng_Li_2023, Chong_Chen_2018, Zheng_2017}. In practice, recommender systems operate under a ranking paradigm, where performance depends on placing the most relevant items in the top-K positions, requiring discriminative representations that clearly separate relevant from non-relevant items. This misalignment between learning objectives and evaluation criteria creates a gap between how semantic signals are modeled and how recommendation quality is assessed. Bridging this gap requires rethinking both the fusion mechanism and the training objective: selectively integrating semantic representations under ranking-driven criteria, and employing contrastive strategies that explicitly promote separation between positive and negative items in the latent space.

In this context, we propose \gls{ghcf}\footnote{Available at: \href{ghcf}{https://github.com/ferreira-eduardo/ghc2f.git}}, which integrates semantic signals extracted from user reviews into the collaborative latent space. Topic distributions derived from review texts are projected and aligned with interaction-based representations, enabling a structured fusion between semantic and collaborative information. To regulate the influence of each signal, the model employs an adaptive gating mechanism that dynamically balances collaborative embeddings and topic-based representations during encoding. Furthermore, \gls{ghcf} is trained under a contrastive, ranking-aware objective that promotes discriminative separation between relevant and non-relevant items. This design addresses a fundamental limitation of purely interaction-driven models: their limited capacity to represent nuanced preference factors, while preserving the compact, expressive latent structure characteristic of autoencoder-based recommenders. To address the identified gaps in ranking optimization and semantic–collaborative alignment, this study makes the following contributions:

\begin{itemize}
    \item A gated hybrid autoencoder framework \gls{ghcf} that integrates topic-based semantic representations into collaborative filtering while preserving compact latent modeling.
    \item An adaptive learning gate that dynamically regulates the contribution of semantic and interaction-based signals, mitigating noise from textual data and improving representation alignment.
    \item A contrastive ranking-aware learning strategy based on the \gls{bpr} loss that explicitly optimizes the separation between relevant and non-relevant items within the fused latent space.
    \item Layer-wise semantic injection, enabling topic information to influence multiple stages of the encoding process rather than only the bottleneck representation.
\end{itemize}

\section{Related Works}
\label{sec:related_works}

This section surveys representative works that inform and contextualize the present study, with emphasis on neural architectures, attention mechanisms, topic modeling, and contrastive strategies for review-aware recommendation. DeepCoNN, proposed by \citet{Zheng_2017}, pioneered the use of parallel \gls{cnn} to learn latent representations of users and items from reviews. While effective in the face of data sparsity, its \gls{cnn}-based architecture struggled with long-range dependencies and cold-start scenarios. Subsequent works introduced more granular modeling approaches: NARRE \cite{Chong_Chen_2018} emphasized the importance of weight reviews, while A3NCF \cite{Zhiyong_Cheng_2028} employed neural topic modeling for aspect-aware representation learning. Similarly, ANR \cite{Chin_2018} utilized aspect-level attention for fine-grained semantic modeling.

Beyond pure text extraction, \citet{Musto_2017} integrated opinion mining and multi-criteria \gls{cf} using sentiment scores to enhance neighborhood calculations. For explainability, CAML \cite{Zhongxia_Chen_2019} used an encoder-selector-decoder architecture with hierarchical co-attention to jointly optimize rating prediction and linguistic explanation. Complementarily, DAML \cite{Liu_2019} combined review-based \gls{cnn} attention with an interaction-based module to capture nonlinear feature relationships. Recent advances leverage structural signals. RGCL \cite{Shuai_2022} incorporates review semantics directly into bipartite graph message-passing via contrastive learning. 

Additionally,~\citet{Shang_2024} further enriched this by integrating hierarchical attention into sentiment-aware neural \gls{cf}, thereby capturing nuanced opinions at the word and sentence levels. More recently, the emergence of \gls{llm} has further reshaped this discussion. ~\citet{Tan2025} question whether specialized review-aware architectures are still necessary and introduce RAREval, a benchmark designed to systematically assess the contribution of reviews under zero-shot, few-shot, and fine-tuning settings. Their results indicate that pretrained LLMs can implicitly capture semantic alignment between users and items, often outperforming traditional review-based architectures, particularly in sparse and cold-start scenarios.

\gls{ghcf} differentiates itself by jointly addressing representation alignment and adaptive fusion between collaborative and review-based signals. It employs a contrastive, ranking-aware objective to explicitly separate relevant from non-relevant items in the shared latent space, while a learned gating mechanism dynamically regulates the contribution of semantic and interaction-based information during encoding. This combination enables controlled integration of heterogeneous signals under a ranking-driven learning framework.

\section{Proposal}

In this section, we present a family of autoencoder-based \gls{cf} models designed to incorporate both interaction data and semantic information derived from user reviews. We begin with the backbone model, denoted as AE-BPR, which leverages an autoencoder architecture optimized with a \gls{bpr} objective to learn latent user–item representations from ratings feedback. Building upon this foundation, we introduce a gated hybrid extension that integrates collaborative signals with semantic representations extracted from textual reviews. The gating mechanism is designed to adaptively balance these complementary sources of information, enabling the model to account for both interaction patterns and content-based features.

Finally, we explore an additional extension that incorporates a contrastive learning objective into the gated hybrid framework. This component aims to further refine the learned representations by encouraging consistency between related samples while distinguishing them from unrelated ones. This progressive formulation allows us to examine the contribution of each component, from the base collaborative model to the hybrid and contrastive extensions, providing a structured basis for the analysis presented in subsequent sections.

\subsection{Autoencoder-based BPR Recommendation (AE-BPR)}

Autoencoder-based \gls{cf} is particularly well-suited to sparse recommendation settings, as it reconstructs user–item interaction vectors via non-linear transformations, enabling the model to capture complex collaborative dependencies beyond linear factorization. Its denoising and compression capabilities make it robust when interaction data is limited or incomplete. However, it operates solely on interaction signals.

\begin{figure}[ht]
    \centering
    \includegraphics[width=1\linewidth]{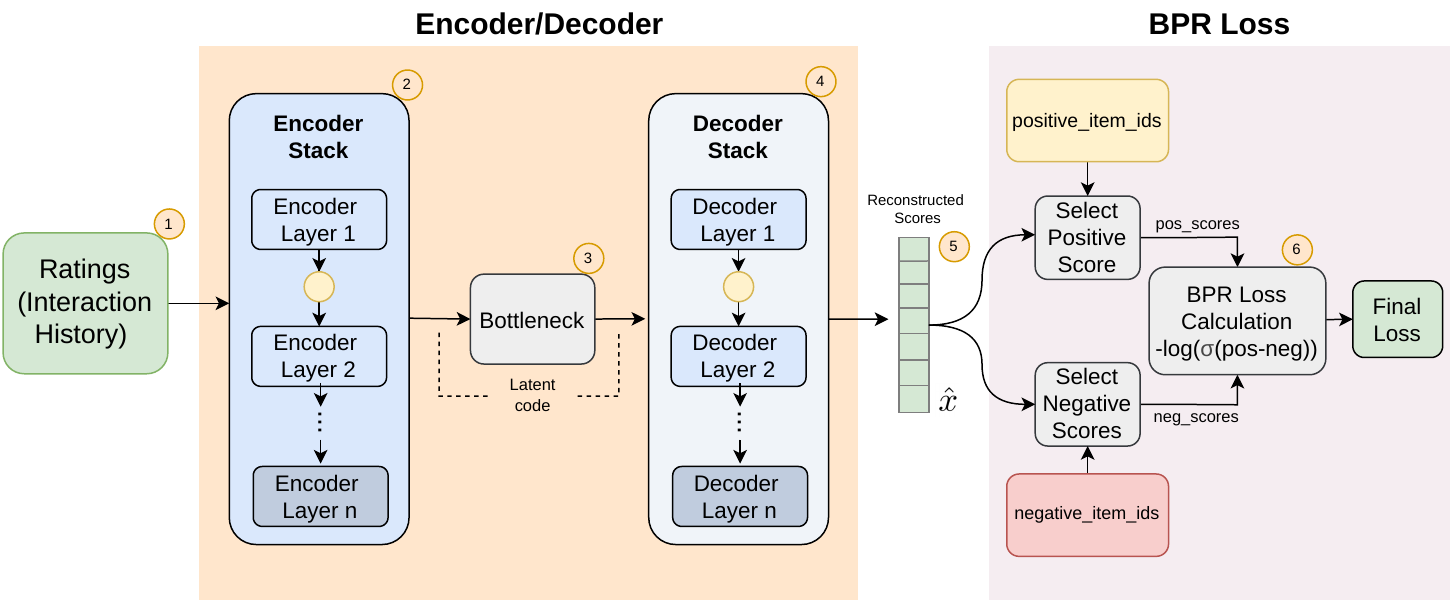}
    \caption{Architecture of the AE-BPR model. User–item interaction histories (1) are processed through an encoder stack (2) to a bottleneck latent code (3). The decoder (4) generates reconstructed scores (5), which are optimized using a pairwise BPR loss (6) based on positive and negative item sampling.}
    \label{fig:ae_bpr}
\end{figure}

The network maps an input vector $x\in \mathcal{R}^n$, representing these interaction profiles Figure~\ref{fig:ae_bpr}(1), to a reconstruction $\hat{x} \in R^n$. The encoder transforms the high-dimensional input into a lower-dimensional latent representation $z \in \mathcal{R}^d$ through a sequence of fully connected layers. Formally, let $h_l$ denote the output of the $l$-th encoder layer, where $h_0=x$. For each layer $l \in {1..k}$, the forward propagation is defined as: 
 \begin{equation}
    h_l =f(W_l h_{l-1} + b_l),    
 \end{equation}
 where $W_l$ and $b_l$ are the weight matrix and bias vector, and $f(\cdot)$ is the non-linear activation function. To prevent overfitting, a dropout layer is applied to the final latent code $z=h_K$ at the bottleneck [Figure~\ref{fig:ae_bpr}(3)].

The decoder stack [Figure~\ref{fig:ae_bpr}(4)] reconstructs the item scores. In our implementation, we support both unconstrained and constrained (tied weights) architectures. In the constrained case, the decoder reuses the encoder's transposed weights, $W_{dec} = W_{enc}^\top$, thereby reducing the number of trainable parameters and providing a regularizing effect. The reconstructed scores $\hat{x}$ [Figure~\ref{fig:ae_bpr}(5)] represent the predicted relevance for all items.

The model is trained using a pairwise \gls{bpr} objective [Figure~\ref{fig:ae_bpr}(6)]. For each user, the loss is calculated by sampling a positive item $i^+$ from the observed interactions and a negative item $i^-$ from the unobserved set. The \gls{bpr} loss encourages the model to assign a higher predicted score $S_{u,i^+}$ to the positive item than the score $S_{u,i^-}$ assigned to the negative item:
\begin{equation}
L_{BPR} = -\ln \sigma(S_{u,i^+} - S_{u,i^-}),
\end{equation}
where $\sigma$ is the sigmoid function. This objective directly optimizes the model for ranking, which is the primary requirement in recommendation tasks. Additionally, to ensure reconstruction accuracy, the model utilizes a \gls{mmse} loss. This restricts error computation to observed entries only, preventing the model from being biased toward the sparsity of the interaction matrix:
\begin{equation}
MMSE = \frac{\sum m_i \cdot (r_i - \hat{x}_i)^2}{\sum m_i}
\end{equation}
where $m_i$ is a binary mask such that $m_i=1$ if an interaction exists ($r_i \neq 0$), and $0$ otherwise.

\subsection{Gated Hybrid Collaborative Filtering for Recommendation}

\begin{figure}[ht]
\centering
\includegraphics[width=1\linewidth]{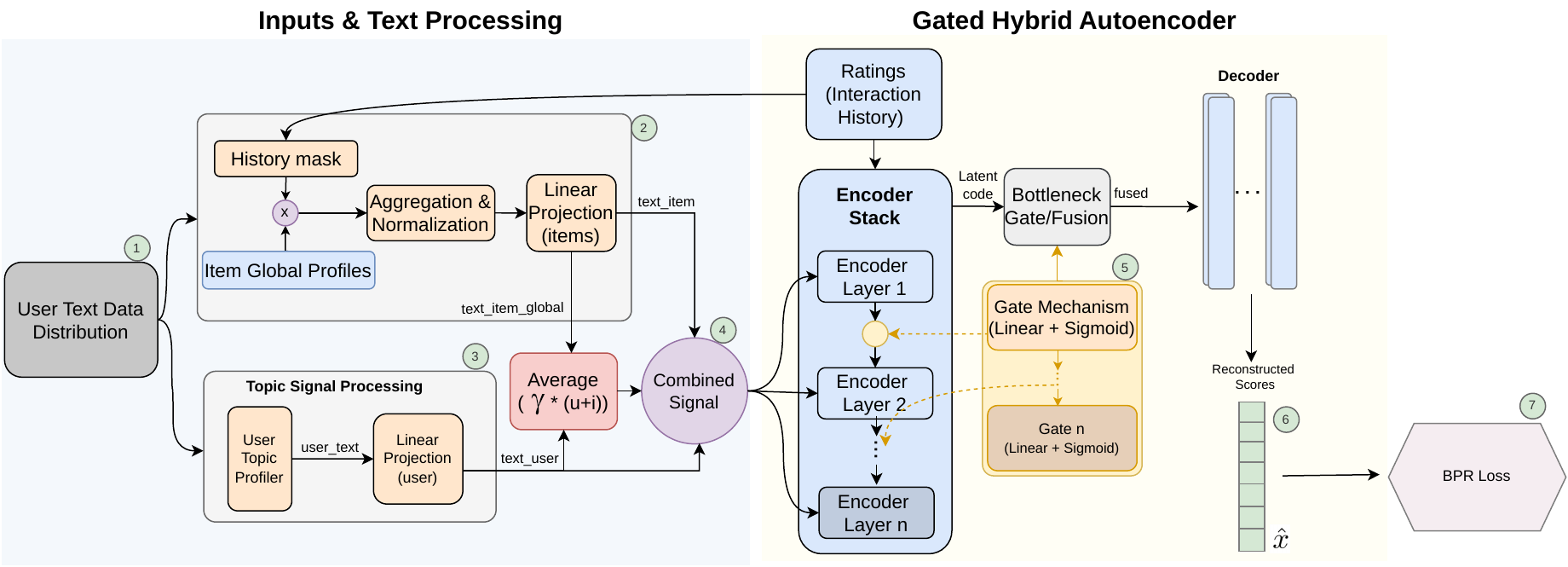}
\caption{End-to-end architecture of the GHCF recommendation model. Semantic signals from user and item distributions (1-4) are integrated into the Autoencoder (5) via a layer-wise gating mechanism to produce reconstructed scores (6) optimized by \gls{bpr} loss (7). The module in hexagon shape is the module reused from Fig~\ref{fig:ae_bpr}(6).}
\label{fig:gated_hybrid}
\end{figure}

To improve the interaction-only models, we propose the \gls{ghcf} architecture. This model incorporates text-based semantic profiles derived from reviews and fuses them with collaborative filtering signals via a layer-wise gating mechanism with a unified end-to-end autoencoder. The complete architecture is illustrated in Figure~\ref{fig:gated_hybrid}.

The process begins by extracting semantic context from two parallel streams. As shown in Figure~\ref{fig:gated_hybrid}(3), the User Text Profiler generates a semantic embedding $z_u$ from the user's text data distribution. This is projected into the collaborative latent space via a linear transformation to obtain the user text signal $t_u$. Simultaneously, the Item Global Profiles are processed. A history mask, derived from the user's interaction history, selects the global text representations of items the user has interacted with. These are aggregated and normalized to form a personalized, item-aware semantic representation, which is then projected to obtain the item-text signal $t_i$. The final combined text signal T is computed as the average of these two components, where the $\gamma$ variable controls the intensity of the combined text:
\begin{equation}
T = \gamma \cdot (t_u + t_i).
\end{equation}

The collaborative pathway uses an autoencoder backbone (Figure~\ref{fig:ae_bpr}) to process the interaction profile $x$. Unlike standard models that fuse information only at the bottleneck, \gls{ghcf} implements a Layer-wise Gated Fusion. For each encoder layer $l$, the collaborative hidden state $h_l$ is fused with the projected text signal $T_l$. A gating mechanism, implemented as a linear transformation followed by a sigmoid activation, computes adaptive coefficients $g_l$ based on the concatenation of both signals:
\begin{equation}
g_l = \sigma(W_{gate} [h_l; T_l] + b_{gate}).
\end{equation}
The fused representation $h_l^{fused}$ is then computed as a weighted combination:
\begin{equation}
h_l^{fused} = g_l \odot h_l + (1 - g_l) \odot T_l,
\end{equation}
this is followed by regularization to mitigate overfitting. This design allows the network to control, on a per-dimension basis, the extent to which semantic information influences collaborative encoding at every level of abstraction.

The decoder stack reconstructs the item scores $\hat{x}$. The model is primarily optimized using a pairwise \gls{bpr} objective. For each batch, the loss encourages the model to assign higher scores to observed interactions ($S_{u,i^+}$) than to sampled negative items ($S_{u,i^-}$):
\begin{equation}
L_{BPR} = -\ln \sigma(S_{u,i^+} - S_{u,i^-}).
\end{equation}

\subsection{Contrastive Extension (GHC2F)}

\begin{figure}[ht]
    \centering
    \includegraphics[width=1\linewidth]{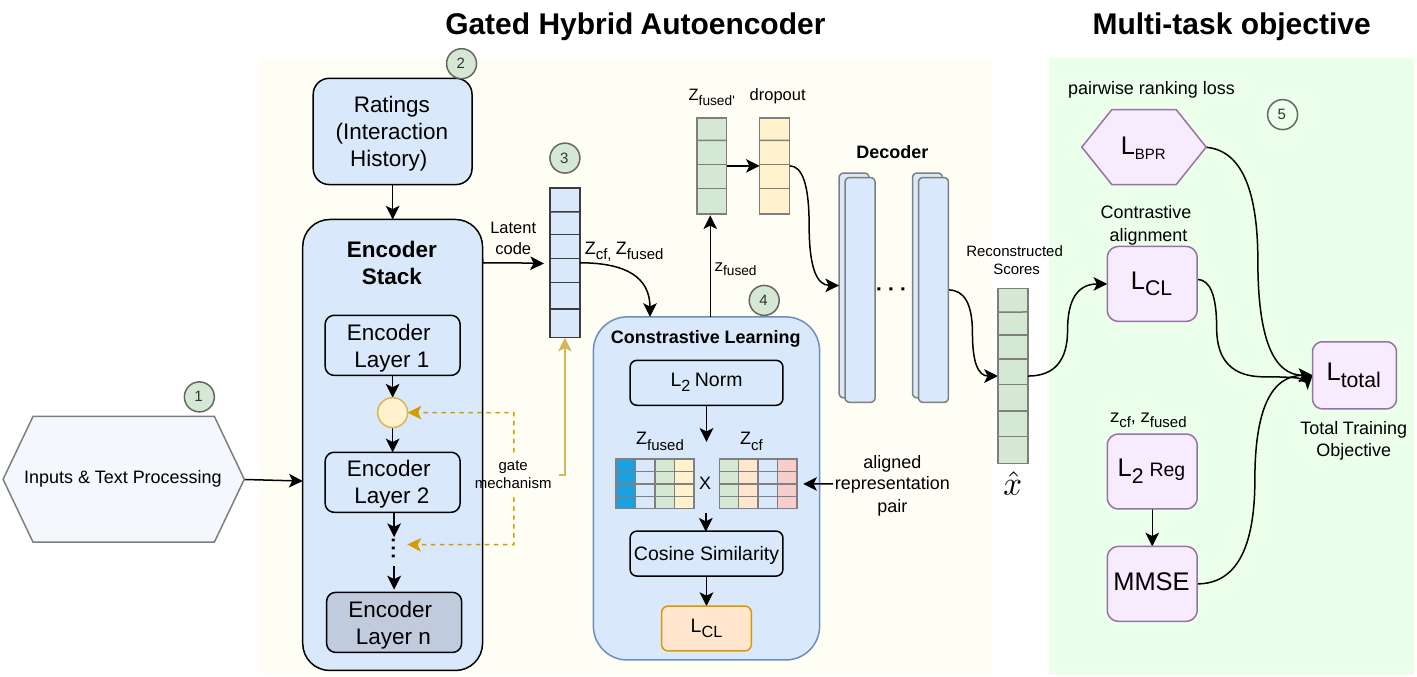}
    \caption{Overview of the proposed gated hybrid collaborative filtering framework. Interaction data and semantic representations are jointly encoded to learn a unified latent representation through a gated fusion mechanism. An optional contrastive objective is incorporated to refine the alignment of representations. The model is trained using a multi-task objective that combines ranking and auxiliary learning signals.}
    \label{fig:ghc2f_arch}
\end{figure}

Building upon the gated architecture, we develop \gls{ghc2f}. This variant leverages contrastive regularization to ensure that user and item embeddings remain discriminative across both collaborative and review-based views. This extension introduces a self-supervised contrastive objective to align the collaborative representation with the semantically enriched latent space learned through gated fusion. The goal is to increase consistency between interaction-driven and hybrid representations, thereby mitigating potential distortions arising from integrating semantic information. An overview of the architecture is presented in Figure~\ref{fig:ghc2f_arch}.

The reused modules are drawn as hexagons, Figure~\ref{fig:ghc2f_arch}(1) from the \gls{ghcf} proposal.  Additionally, the modules in Figure~\ref{fig:ghc2f_arch}(2,3) follow the pattern to measure in the contrastive model in Figure~\ref{fig:ghc2f_arch}(4).

The core innovation of \gls{ghc2f} is centered on its dual-pathway bottleneck representation [see Figure~\ref{fig:ghc2f_arch}(4)]. In this architecture, the encoder generates two complementary latent vectors for each user $u$: a collaborative representation ($z_{cf}$), derived solely from interaction-driven signals, and a fused representation ($z_{fused}$), which is applied after the gated mechanism integrates collaborative and semantic signals. This bifurcation allows the model to explicitly align raw collaborative patterns with review-enriched features.

To promote consistency between these two views, we adopt a contrastive learning objective based on the InfoNCE formulation~\cite{infoce_2019}. Before computing the contrastive loss, an alignment function (e.g., a linear projection) maps $z_{cf}$ to the same latent space as $z_{fused}$. Both representations are then $L_2$-normalized:

\begin{equation}
\bar{z}_{cf} = \frac{z_{cf}}{\|z_{cf}\|_2}, 
\quad 
\bar{z}_{fused} = \frac{z_{fused}}{\|z_{fused}\|_2}.
\end{equation}

Given a mini-batch $B$, the pair $(\bar{z}_{cf,u}, \bar{z}_{fused,u})$ is treated as a positive pair [Figure~\ref{fig:ghc2f_arch}(5)], while representations from different users form negative pairs. The contrastive objective is defined as:

\begin{equation}
L_{CL} = - \sum_{u \in B} 
\log 
\frac{
\exp\left( \text{sim}(\bar{z}_{cf,u}, \bar{z}_{fused,u}) / \tau \right)
}{
\sum_{v \in B} 
\exp\left( \text{sim}(\bar{z}_{cf,u}, \bar{z}_{fused,v}) / \tau \right)
}
\label{eq:cl_loss}
\end{equation}

where $\text{sim}(\cdot)$ denotes cosine similarity and $\tau$ is a temperature hyperparameter.

The model is trained end-to-end using a multi-task objective that combines ranking, contrastive alignment, and regularization terms:

\begin{equation}
L_{total} = L_{BPR} + \lambda_{CL} L_{CL} + \lambda_{reg_w} L_{reg_w} + \lambda_{reg_i} L_{reg_i}.
\end{equation}

Here, $L_{BPR}$ is the primary pairwise ranking objective, $L_{CL}$ is the contrastive loss defined in Equation~\ref{eq:cl_loss}, and $L_{reg_w}$ and $L_{reg_i}$ correspond to $L_2$ regularization on model parameters and input representations, respectively. The coefficients $\lambda$ control the contribution of each term.

Overall, GHC2F aims to learn latent representations that remain discriminative for ranking while maintaining alignment between collaborative and hybrid semantic spaces.

\section{Experimental Evaluation}
\label{sec:experimental_evaluation}

\subsection{Methodology}

This study employs a comparative experimental design to evaluate whether a gated hybrid autoencoder improves ranking performance in review-based recommendation settings. The proposed model extends a collaborative autoencoder optimized with a pairwise \gls{bpr} objective by introducing contrastive alignment between rating-derived collaborative codes and review-derived topic representations. A learnable gating mechanism controls the amount of textual information injected into the intermediate latent layers. 

Experiments are conducted on three benchmark datasets, Amazon Movies \& TV, IMDb, and Rotten Tomatoes, which provide explicit ratings and associated textual reviews, enabling the evaluation of hybrid recommendation strategies that combine behavioral and content-based signals. A \gls{loo} evaluation protocol is employed, in which the most recent interaction for each user is reserved for testing and the second most recent for validation. To construct pairwise training instances, positive and negative items are defined using a user-specific z-score normalization over ratings, which accounts for heterogeneous user rating behavior. Items with positive standardized scores are treated as positives, while negatives include both unseen items and items rated below the user’s mean (z-score $< 0$), enlarging the negative pool with informative ``disliked'' observations in addition to unobserved candidates.

The proposed approach is compared against several state-of-the-art review-aware recommendation models. Specifically, CARL, ConvMF, DAML, DeepCoNN, and TARMF were implemented following the unified experimental framework proposed by \citet{Bittencourt_2025}, ensuring consistency in preprocessing, training, and evaluation procedures. These baselines are complemented by GRAFT, KANN, and LETTER, enabling a comprehensive comparison with recent review-aware and hybrid recommendation methods. Model performance is assessed using standard top-N ranking metrics, including \gls{hr}, \gls{ndcg}, and \gls{mrr}. These metrics evaluate the model’s ability to correctly rank relevant items within a candidate set. Statistical significance tests are conducted to validate observed performance differences across models.

\subsection{Datasets}

\begin{table*}[ht]
    \centering
    \caption{Comparative statistics of the datasets before and after preprocessing.}
    \resizebox{\textwidth}{!}{
    \begin{tabular}{l*{6}{c}}
        \toprule
        & \multicolumn{2}{c}{\textbf{Amazon Movies and TV}}  & \multicolumn{2}{c}{\textbf{IMDb}}  & \multicolumn{2}{c}{\textbf{Rotten Tomatoes}} \\
        \cmidrule(lr){2-3}\cmidrule(lr){4-5}\cmidrule(lr){6-7}
        \textbf{Metric} & \textbf{Raw} & \textbf{Processed}& \textbf{Raw} & \textbf{Processed}& \textbf{Raw} & \textbf{Processed} \\
        \midrule
        Reviews         & 17.3M     & 1.4M      & 932.4K    & 264.9K    & 1.1M      & 543.5K   \\
        Users           & 6.5M      & 76K       & 427K      & 7k        & 11K       & 2.6k     \\
        Items           & 747.7K    & 27.3K     & 1.1K      & 1.1K      & 17.7K     & 17.4K    \\
        Reviews/User    & 2.664     & 19.12     & 3.28      & 37.07     & 100.06    & 208.17   \\
        Reviews/Item    & 23.173    & 53.18     & 12.28     & 230.40    & 63.77     & 31.13    \\
        Sparsity        & 99.99\%   & 99.93\%   & 99.81\%   & 96.77\%   & 99.42\%   & 98.80\%  \\
        \bottomrule
    \end{tabular}
    }
    \label{tab:datasets}    
\end{table*}

By combining Amazon Movie and TV, IMDb, and Rotten Tomatoes, we gain access to both professional and user perspectives, enriching our understanding of recommendation systems research. Amazon and IMDb offer detailed user-generated reviews, enabling a deeper understanding of individual preferences. Additionally, including Rotten Tomatoes is especially valuable because it uniquely provides both professional critic reviews and aggregated user ratings and scores. Combining nuanced user opinions with expert critiques provides a comprehensive foundation for informed decision-making. 

After cleaning and filtering, the summary statistics of the pre-processed datasets are presented in Table~\ref{tab:datasets}. All datasets were processed to remove empty or null entries and to retain only users with at least 10 interactions. Light cleaning was applied during textual preprocessing to remove specific characters, hyperlinks, and other residual noise from each dataset. For example, in the Amazon dataset, reviews that start with ``...'' or ``!!!'' are removed. Furthermore, we select only English reviews. Finally, small reviews like ``Great movie'', ``Excellent'', or ``Terrible'' were also maintained, as were the emojis, case, and punctuation. The idea is to keep all the expressions for the original datasets.  The specific preprocessing steps applied to each dataset, alongside the resulting statistics, are detailed in the following subsections.

\subsection{Evaluation Metrics}

Given that the proposed model is optimized using \gls{bpr}, which directly models pairwise preferences in implicit-feedback settings, evaluation is conducted using ranking-based metrics rather than pointwise rating-accuracy measures. Following the \gls{loo} protocol, for each user, one ground-truth positive interaction is ranked against 99 sampled negative items. The relative position of the positive item within this candidate set determines the evaluation outcomes. Let rank\_u denote the rank position of the held-out positive item for user u.~\gls{hr} measures the proportion of users for whom the ground-truth item appears within the top-K positions. 
\begin{equation}
    HR@K = \frac{1}{|U|} \sum I (rank \leq K),
\end{equation}
where I($\cdot$) denotes the indicator function.~\gls{hr} captures the model's ability to retrieve relevant items among the most exposed recommendations.

Additionally, users predominantly focus on the highest-ranked recommendations, and evaluation prioritizes ranking effectiveness in top positions over overall retrieval coverage. Accordingly, we employ \gls{mrr} and \gls{ndcg} to assess the quality of item ordering. 
~\gls{ndcg} evaluates ranking performance within the top-K positions by applying logarithmic discounting to lower ranks, capturing the diminishing utility associated with deeper positions in the list.  Together, these metrics provide a position-sensitive assessment of recommendation quality, aligning with practical user interaction patterns and the pairwise ranking objective optimized during training.   
\begin{equation}
    nDCG@K = \frac{1}{|U|} \sum_{u\in U}
    \begin{cases}
        \frac{1}{log_2(rank_u +1)} & \text{if } rank_u \leq K\\
        0 & otherwise
    \end{cases}.
\end{equation}
\gls{mrr} computes the average reciprocal position of the first relevant item, thereby emphasizing early precision and strongly penalizing relevant items placed lower in the ranking,
\begin{equation}
    MRR = \frac{1}{|U|}\sum_{u\in U}\frac{1}{rank_u}.
\end{equation}

\subsection{Baselines}

Despite the rapid development of deep learning techniques, autoencoder methods remain strong and competitive baselines in many recommendation tasks. To provide a comprehensive evaluation, we compare our approach with eight baseline models from different methodological families. These include matrix factorization–based methods, models that incorporate review text information, architectures that employ attention mechanisms to capture user–item interactions, models that use \gls{cnn} to extract features from reviews, and autoencoder-based approaches. This selection covers both classical and neural architectures commonly used in review-aware recommendation scenarios.

For the baselines, we employ the framework proposed in~\cite{irev_2025} to evaluate the models over the 2014–2024 period, and supplement it with LETTER, KANN, and GRAFT, which were run directly from their respective public repositories. All models are evaluated using the hyperparameter configurations specified in their original publications, without further tuning, to ensure a fair and reproducible comparison. To control for differences in input representations, all models share the same text encoder, \texttt{all-mpnet-base-v2}, providing a consistent semantic embedding space across methods.

\begin{itemize}
    \item CARL. A context-aware representation learning model that combines review-based and interaction-based features using Factorization Machines to model higher-order user–item interactions \cite{Libing_Wu_2019}.
    \item ConvMF. A CNN-based review-aware recommender that extracts user and item representations from aggregated review texts and integrates them within a matrix factorization framework \cite{Kim_2016}.
    \item DAML. A dual-attention model that jointly learns rating and review features, integrating them via neural factorization machines to capture nonlinear feature interactions \cite{Liu_2019}.
    \item DeepCoNN. A dual-network architecture that independently learns user and item embeddings from reviews and couples them through a shared interaction layer \cite{Zheng_2017}.
    \item GRAFT. A graph attention-based recommender that embeds review-derived topic representations into a user–item graph with adaptive gating \cite{Eduardo_2025}.
    \item KANN. A knowledge graph-enhanced recommender that models internal review attention and external user–item relations to learn structured preference representations \cite{Liu_2023}.
    \item LETTER. A GNN-based review-aware model that leverages homogeneous user and item graphs to learn sentiment-informed representations \cite{Jiwon_Son_2025}.
    \item TARMF. A hybrid model combining matrix factorization with an attention-based GRU to integrate rating data and sequential review representations \cite{Yichao_Lu_2018}.
\end{itemize}

\subsection{Topic Extraction}

The objective of the topic extraction stage is to uncover nuanced aspects of user experience expressed in reviews, such as emotional impact, immersion, or technical qualities, that are not captured by standard metadata. The datasets differ substantially in review length, ranging from very short Rotten Tomatoes reviews to long, descriptive IMDb reviews, with Amazon reviews falling in the middle. 

To address this variability, we evaluated two sentence embedding models during hyperparameter search: a lightweight transformer \texttt{(thenlper/gte-small)}~\cite{li2023towards}, which offers efficient inference and scalability, and a larger, high-performance model \texttt{(all-mpnet-base-v2)}\footnote{Available at: \href{sentence-transformers/all-mpnet-base-v2}{https://huggingface.co/sentence-transformers/all-mpnet-base-v2}}, which produces richer semantic representations at a higher computational cost. 

This comparison enables an analysis of the trade-off between topic quality and efficiency. For topic discovery, high-dimensional embeddings are first reduced using Uniform Manifold Approximation and Projection (UMAP) and then clustered with K-Means to group semantically similar reviews, prioritizing stable cluster assignments over explicit outlier detection. Topic representation follows the BERTopic framework~\cite{grootendorst2022} and uses a KeyBERT-inspired approach to extract representative keywords aligned with document embeddings. Domain-specific stopwords are removed, and n-grams up to trigrams are considered to improve semantic expressiveness. The resulting pipeline produces interpretable topic labels, per-review topic distributions, and low-dimensional visualizations, supporting subsequent analysis and model integration.

\subsection{Hyperparameters Sensitivity}

We conducted a hyperparameter sensitivity analysis using Optuna\footnote{Available at: \href{Optuna}{https://optuna.org/}}, with 20 trials per dataset, on a user-stratified 20\% subsample to reduce computational cost while preserving the recommendation structure. Each configuration was trained under the leave-one-out 
protocol (fold 0) for 15 epochs, optimizing the validation BPR ranking loss. 
The search space included dropout \{0.2, 0.5\}, learning rate \{1e$^{-4}$, 1e$^{-3}$\}, activation function \{selu, relu\}, and three encoder 
architectures {total\_items, 4096\}, \{total\_items, 4096, 2048\}, \{total\_items, 4096, 2048, 1024\} to evaluate the impact of model capacity. Topic latent dimensionality was fixed to ensure comparability across trials. The best configuration for each dataset was selected based on the minimum validation loss and hit rate, and subsequently used in the final 5-fold comparative evaluation. The best set of hyperparameters is composed by layers \{total\_items, 4096\}, learning rate 1e$^{-4}$, and \textit{selu} as activation function.

\section{Results}

\subsection{Topic Extraction}

\begin{table}[ht]
\centering
\caption{Representative topics extracted using the BERTopic model across the IMDb, Amazon Movies and TV, and Rotten Tomatoes datasets, where ``T'' means Topic. Each column lists the most salient keywords for each identified topic, highlighting the dominant semantic patterns in user reviews.}
\label{tab:all_topics}
    \begin{tabular}{l p{3.4cm} p{3.8cm} p{3.8cm}}
    \toprule
    \textbf{T} & \textbf{IMDb} & \textbf{Amazon Movies and TV} & \textbf{Rotten Tomatoes} \\
    \midrule
    1 & good story really time & story line disappointed character storyline & characters story plot action \\
    2 & love life music story & saw bad quality recommend & director performances music story \\
    3 & batman man marvel superhero & story line comedy recommend character & fun good entertaining funny \\
    4 & alien action good time & character characters plot comedy & documentary war story life \\
    5 & tarantino good time action & special effects characters saw book & performance character best actor \\
    6 & animation story disney animated & story line entertaining storyline lost & horror thriller scary genre \\
    7 & war story soldiers time & recommend documentary worth book & love romantic comedy story \\
    8 & horror dead zombie good & recommend storyline entertaining lots action & comedy funny laughs jokes \\
    9 & story good time best & recommend entertaining favorite bad & kids disney animated animation \\
    10 & & downton downton abbey recommend highly & sequel original action remake \\
    11 & & terminator trilogy star wars saw & bond eastwood action western \\
    12 & & transformers special effects book characters & \\
    13 & & quality special effects harry worth & \\
    \bottomrule
    \end{tabular}
\end{table}

Initially, we employed the coherence metric to screen multiple model configurations, thereby efficiently narrowing the search space~\cite{coherence}.  Recognizing that statistical coherence does not always align with human-readable themes, we employed an iterative refinement process to filter out noisy clusters and verify that each candidate topic represented a distinct, interpretable concept in the review corpus.
This process involved examining top terms, representative reviews, and inter-topic overlap to identify and address issues such as spurious, fragmented, or overly generic themes. When necessary, topics were merged, split, or relabeled. The resulting set of topics achieves a balance between statistical consistency and semantic interpretability, serving as the foundation for subsequent analyses.

We hypothesize that reviews encode aspects not present in basic metadata (e.g., sound quality, narrative structure, themes). For each corpus, we fit \(K=15\) topics and then applied a qualitative screen: topics with prevalence \(<10\%\) of documents were removed or merged, and entity-centric or overly specific themes (e.g., actor or company names) were filtered to favor generalizable aspects. The final inventories contain 13 topics for IMDb, 11 for Amazon Movies \& TV, and 9 for Rotten Tomatoes; Table~\ref{tab:all_topics} reports the resulting sets. 

Table~\ref{tab:all_topics} displays two stable patterns across datasets: a narrative core (comprising story, plot, and character) and recurring genre clusters (horror, comedy, war/documentary, animation/Disney, and action/special effects). These patterns support the hypothesis that reviews surface non-trivial aspects beyond basic metadata, including production attributes (e.g., special effects), craft signals (e.g., director, performances, and music), and evaluative lines (e.g., quality and recommendation). Dataset profiles vary by context and text length: IMDb’s longer reviews yield more detailed, title- or franchise-level topics (e.g., Batman/Marvel, Alien, Tarantino, Terminator/Star Wars/Saw, Transformers, Harry Potter, Downton Abbey). 

Although our selection criteria avoid overly specific entities such as actor or company names, we retained these franchise topics because they are coherent and frequent. Amazon, collected in a purchasing context, emphasizes evaluation and product orientation (recommend, quality, sequel/original/remake, book), reflecting concerns about continuity, format, and buyer satisfaction. Rotten Tomatoes, based on short blurbs, focuses on craft descriptors (director, performances, best actor, music) and broad genres rather than title-specific synopses. The table also illustrates context-driven ambiguity: ``saw'' is a verb in Amazon but denotes a horror franchise in IMDb. Frequent generic intensifiers (good, best, really, time) express sentiment but add little topical detail. In summary, the topics capture narrative and genre, along with production and evaluation signals, while revealing dataset-specific fingerprints: franchise focus on IMDb, product evaluation on Amazon, and craft-oriented critique on Rotten Tomatoes.

\subsection{Recommendation Ranking}

Table~\ref{tab:results_summary} presents the top-10 ranking performance across the three datasets, revealing consistent and substantial improvements of the proposed \gls{ghcf} model over existing review-aware baselines, particularly in Amazon and Rotten Tomatoes. On the Amazon Movies \& TV dataset, \gls{ghcf} achieves the highest \gls{hr}@10 (0.2710) and \gls{ndcg}@10 (0.1621), outperforming strong baselines such as TARMF and DAML by a clear margin. The improvement in \gls{hr} indicates that the model more frequently places relevant items within the Top-10 list, while the gain in \gls{ndcg} reflects better positional ordering of relevant items. Although LETTER achieves the highest \gls{mrr} (0.1955), \gls{ghcf} remains competitive (0.1459), suggesting that while LETTER may place a single relevant item earlier in some cases, \gls{ghcf} demonstrates stronger overall ranking consistency across the Top-10.

On IMDb, \gls{ghcf} again attains the best \gls{hr}@10 (0.2187), demonstrating superior ability to retrieve relevant items. However, LETTER achieves the highest \gls{ndcg} (0.1443) and \gls{mrr} (0.1931), indicating stronger early-ranking precision. This suggests that while \gls{ghcf} improves overall retrieval coverage, LETTER may prioritize top-position refinement more aggressively on this dataset. Nonetheless, the substantial \gls{hr} gain confirms the effectiveness of contrastive rating–review alignment in improving candidate retrieval.

The most striking results appear on Rotten Tomatoes, where \gls{ghcf} significantly outperforms all baselines in \gls{hr}@10 (0.6352) and \gls{ndcg}@10 (0.4228), nearly tripling the \gls{hr} of most competing models. This indicates that the proposed gating and contrastive alignment mechanisms are particularly effective in datasets where review semantics play a stronger role in user preference modeling. Although LETTER achieves the highest \gls{mrr} (0.3805), \gls{ghcf} remains highly competitive (0.3729) while delivering substantially better overall ranking depth.
Across datasets, two consistent patterns are evident. First, models that rely solely on collaborative signals or shallow text integration (e.g., ConvMF, DeepCoNN) underperform compared to architectures that more tightly couple reviews and interaction signals. Second, the proposed method shows strong gains in \gls{hr} and \gls{ndcg}, indicating improved ranking robustness and a better distribution of relevant items within the recommendation list. 

The contrastive alignment between rating-derived collaborative codes and review-derived topic representations, combined with adaptive layer-wise gating, appears to enhance representation coherence without sacrificing ranking stability. Overall, the results support the hypothesis that structured, contrastive integration of ratings and reviews yields measurable improvements in Top-N recommendation performance, particularly in datasets with rich textual signals.

\begin{table*}[ht]
\centering
\caption{Top-10 ranking performance across datasets, evaluated using ranking metrics. Results are averaged over 5-fold cross-validation. The best results for each dataset and metric are highlighted in bold, and the seconds are underlined.}
\resizebox{\textwidth}{!}{
\begin{tabular}{llccc}
\toprule
\textbf{Dataset} & \textbf{Model} & \textbf{HR@10} & \textbf{NDCG@10} & \textbf{MRR} \\
\midrule
\multirow{13}{*}{Amazon}
& CARL & 0.1763 ($\pm$ 0.0191) & 0.0955 ($\pm$ 0.0121) & 0.0915 ($\pm$ 0.0098) \\
& ConvMF & 0.1663 ($\pm$ 0.0357) & 0.0924 ($\pm$ 0.0299) & 0.0906 ($\pm$ 0.0269) \\
& DAML & 0.1792 ($\pm$ 0.0188) & 0.0998 ($\pm$ 0.0095) & 0.0964 ($\pm$ 0.0074) \\
& DeepCoNN & 0.1534 ($\pm$ 0.0448) & 0.0841 ($\pm$ 0.0358) & 0.0851 ($\pm$ 0.0314) \\
& GRAFT & 0.0884 ($\pm$ 0.0019) & 0.0400 ($\pm$ 0.0014) & 0.0256 ($\pm$ 0.0012) \\
& KANN & 0.1069 ($\pm$ 0.0057) & 0.0491 ($\pm$ 0.0028) & 0.0317 ($\pm$ 0.0020) \\
& LETTER & 0.0801 ($\pm$ 0.0137) & 0.0948 ($\pm$ 0.0171) & \textbf{0.1955} ($\pm$ 0.0357) \\
& TARMF & 0.2307 ($\pm$ 0.0518) & 0.1431 ($\pm$ 0.0382) & 0.1358 ($\pm$ 0.0331) \\
& AE\_BPR \textit{(ours)} & \textbf{0.3018} ($\pm$ 0.0100) & \textbf{0.1857} ($\pm$ 0.0074) & \underline{0.1665} ($\pm$ 0.0064) \\
& GHC2F\_Text \textit{(ours)} & 0.1742 ($\pm$ 0.0084) & 0.0988 ($\pm$ 0.0050) & 0.0976 ($\pm$ 0.0040) \\
& GHC2F\_Topic \textit{(ours)} & 0.2610 ($\pm$ 0.0104) & 0.1526 ($\pm$ 0.0088) & 0.1381 ($\pm$ 0.0080) \\
& GHCF\_Text \textit{(ours)} & 0.2589 ($\pm$ 0.0098) & 0.1549 ($\pm$ 0.0082) & 0.1400 ($\pm$ 0.0074) \\
& GHCF\_Topic \textit{(ours)} & \underline{0.2710} ($\pm$ 0.0078) & \underline{0.1621} ($\pm$ 0.0066) & 0.1459 ($\pm$ 0.0061) \\

\midrule
\multirow{13}{*}{IMDb}
& CARL & \underline{0.1725} ($\pm$ 0.0042) & 0.0803 ($\pm$ 0.0012) & 0.0767 ($\pm$ 0.0009) \\
& ConvMF & 0.1423 ($\pm$ 0.0100) & 0.0651 ($\pm$ 0.0041) & 0.0656 ($\pm$ 0.0017) \\
& DAML & 0.1300 ($\pm$ 0.0058) & 0.0596 ($\pm$ 0.0030) & 0.0625 ($\pm$ 0.0022) \\
& DeepCoNN & 0.1376 ($\pm$ 0.0074) & 0.0638 ($\pm$ 0.0038) & 0.0653 ($\pm$ 0.0028) \\
& GRAFT & 0.1189 ($\pm$ 0.0110) & 0.0606 ($\pm$ 0.0070) & 0.0377 ($\pm$ 0.0046) \\
& KANN & 0.0908 ($\pm$ 0.0127) & 0.0447 ($\pm$ 0.0059) & 0.0272 ($\pm$ 0.0039) \\
& LETTER & 0.1640 ($\pm$ 0.0041) & \textbf{0.1443} ($\pm$ 0.0021) & \textbf{0.1931} ($\pm$ 0.0043) \\
& TARMF & 0.1415 ($\pm$ 0.0110) & 0.0653 ($\pm$ 0.0051) & 0.0668 ($\pm$ 0.0047) \\
& AE\_BPR \textit{(ours)} & 0.1643 ($\pm$ 0.0139) & 0.0893 ($\pm$ 0.0071) & 0.0899 ($\pm$ 0.0063) \\
& GHC2F\_Text \textit{(ours)}  & 0.1252 ($\pm$ 0.0080) & 0.0679 ($\pm$ 0.0031) & 0.0730 ($\pm$ 0.0018) \\
& GHC2F\_Topic \textit{(ours)}  & 0.1557 ($\pm$ 0.0091) & 0.0844 ($\pm$ 0.0044) & 0.0855 ($\pm$ 0.0034) \\
& GHCF\_Text \textit{(ours)} & 0.1601 ($\pm$ 0.0137) & 0.0859 ($\pm$ 0.0073) & 0.0867 ($\pm$ 0.0059) \\
& GHCF\_Topic \textit{(ours)} & \textbf{0.2187} ($\pm$ 0.0092) & \underline{0.1174} ($\pm$ 0.0060) & \underline{0.1099} ($\pm$ 0.0049) \\

\midrule
\multirow{13}{*}{Rotten Tomatoes}
& CARL & 0.2127 ($\pm$ 0.0058) & 0.1084 ($\pm$ 0.0054) & 0.0995 ($\pm$ 0.0061) \\
& ConvMF & 0.1780 ($\pm$ 0.0034) & 0.0900 ($\pm$ 0.0017) & 0.0864 ($\pm$ 0.0017) \\
& DAML & 0.2028 ($\pm$ 0.0160) & 0.1014 ($\pm$ 0.0079) & 0.0934 ($\pm$ 0.0059) \\
& DeepCoNN & 0.1745 ($\pm$ 0.0093) & 0.0871 ($\pm$ 0.0039) & 0.0838 ($\pm$ 0.0024) \\
& GRAFT & 0.2026 ($\pm$ 0.0074) & 0.0894 ($\pm$ 0.0041) & 0.0555 ($\pm$ 0.0039) \\
& KANN & 0.1022 ($\pm$ 0.0171) & 0.0478 ($\pm$ 0.0098) & 0.0311 ($\pm$ 0.0074) \\
& LETTER & 0.1872 ($\pm$ 0.0220) & 0.2965 ($\pm$ 0.0467) & \textbf{0.3805} ($\pm$ 0.0726) \\
& TARMF & 0.1906 ($\pm$ 0.0155) & 0.0989 ($\pm$ 0.0120) & 0.0943 ($\pm$ 0.0103) \\
& AE\_BPR \textit{(ours)} & 0.4368 ($\pm$ 0.0841) & 0.2321 ($\pm$ 0.0450) & 0.1912 ($\pm$ 0.0303) \\
& GHC2F\_Text \textit{(ours)} & 0.2614 ($\pm$ 0.0135) & 0.1445 ($\pm$ 0.0126) & 0.1319 ($\pm$ 0.0122) \\
& GHC2F\_Topic \textit{(ours)} & \underline{0.5355} ($\pm$ 0.0503) & \underline{0.3829} ($\pm$ 0.0432) & 0.3534 ($\pm$ 0.0395) \\
& GHCF\_Text \textit{(ours)} & 0.3540 ($\pm$ 0.0086) & 0.1905 ($\pm$ 0.0057) & 0.1646 ($\pm$ 0.0053) \\
& GHCF\_Topic \textit{(ours)} & \textbf{0.6352} ($\pm$ 0.0222)& \textbf{0.4228} ($\pm$ 0.0204) & \underline{0.3729} ($\pm$ 0.0197) \\
\bottomrule
\end{tabular}
}
\label{tab:results_summary}
\end{table*}

\subsection{Statistical Analysis}

To identify the top-performing model across diverse scenarios, we move beyond raw performance scores to evaluate ranking consistency across all metrics and datasets. Following the statistical framework established by \citet{Demsar_2006}, we adopted a methodology based on Global Average Rankings and non-parametric hypothesis testing. The objective workflow to determine algorithmic superiority was structured as follows:
\begin{enumerate}
    \renewcommand{\labelenumi}{\roman{enumi}.}
    \item Multi-Metric Consolidation via \gls{hv}: To unify the multi-dimensional evaluation (\gls{hr}, \gls{ndcg}, and \gls{mrr}), we employed the \gls{hv} Indicator with a reference point at [0, 0, 0]. This approach maps the multi-objective metric vector into a single scalar that strictly respects Pareto dominance, providing a robust input for subsequent statistical analysis.
    \item Context-Specific Ranking: For each dataset-metric combination, algorithms were ranked from 1 (best) to k (worst). This procedure addresses the issue of incommensurability, ensuring that differing performance scales across domains do not bias the global comparison.
    \item Friedman Test for Global Significance: We applied the Friedman test to determine whether the observed differences in rankings across all contexts are statistically significant ($ p < 0.05$). This step ensures that an algorithm’s performance is not due to stochastic fluctuations.
    \item Global Average Rank and Post-hoc Analysis: Finally, we calculated the average rank for each model across all scenarios. The algorithm with the lowest Global Average Rank is identified as the most consistent and balanced performer. To visualize significant differences between specific pairs of algorithms, we used \gls{cd} diagrams based on the Nemenyi post hoc test.
    \item By integrating the Hypervolume Indicator with Demšar's statistical framework, we ensure that the identified best model achieves a superior and statistically significant balance across all recommendation dimensions.
\end{enumerate}

\subsection{Discussion}
\label{sec:discussion}

Figure~\ref{fig:matrix_nemenyi} shows the performance of the nine algorithms across all datasets and folds, revealing a clear, consistent hierarchical structure. \gls{ghcf} achieves the best overall results, maintaining top positions across nearly all folds and datasets, indicating strong effectiveness and robustness. LETTER and CARL follow as competitive alternatives, with relatively stable rankings and limited fluctuation. In contrast, methods such as GRAFT and KANN consistently rank lowest, reflecting weaker performance and reduced competitiveness across scenarios. Intermediate models (e.g., TAR-MF, DAML, ConvMF, and DeepCoNN) exhibit moderate performance and lower stability, occasionally improving on specific folds but lacking consistent dominance. 

\begin{figure}
    \centering    \includegraphics[width=1\linewidth]{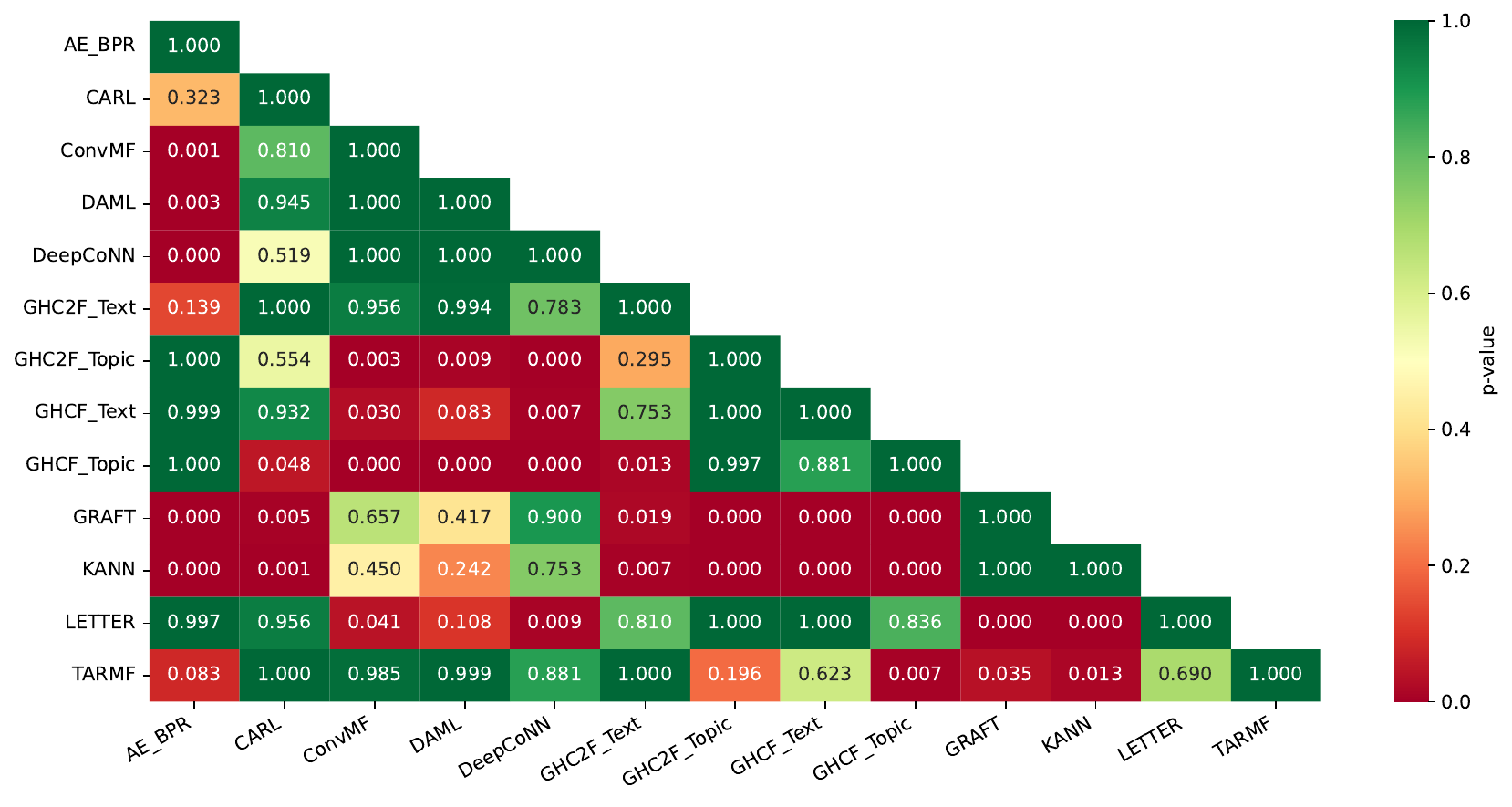}
    \caption{Post-hoc Nemenyi test results showing the pairwise comparison of $p$-values across nine algorithms. The heatmap indicates the statistical significance of performance differences, with values at $p < 0.05$ (red cells) denoting significant differences between models. GHC2F demonstrates superior performance, showing significant gains over GRAFT, KANN, and DeepCoNN.}
    \label{fig:matrix_nemenyi}
\end{figure}

Overall, the results indicate that the leading methods not only achieve superior average performance but also maintain stability across different data partitions, reinforcing their superiority in the comparative analysis.
Complementary, Figure~\ref{fig:diagram_cd} shows the \gls{cd} diagram, which reinforces the hierarchical pattern observed in the previous analysis by incorporating statistical significance into the comparison. \gls{ghcf} is confirming its superior average ranking, while LETTER and CARL appear close behind, forming a leading group. The connecting bars indicate that several of these top-performing methods do not differ statistically, suggesting that although \gls{ghcf} achieves the best mean rank, its advantage over the immediate competitors is not consistently large enough to be considered statistically distinct across all datasets. In contrast, models positioned toward the right, particularly GRAFT and KANN, are clearly separated from the leading group, indicating significantly weaker performance. 

\begin{figure}
    \centering
\includegraphics[width=1\linewidth]{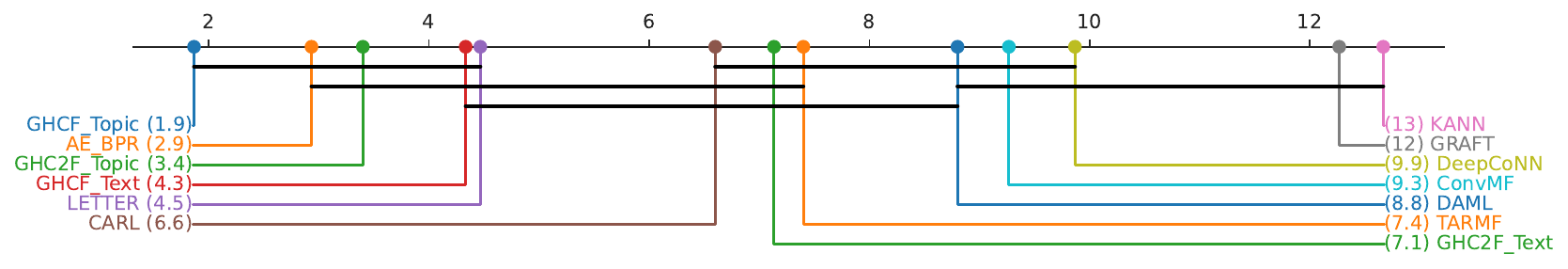}
    \caption{ Average Ranking Comparison. Models further to the left perform better. Connecting bars indicate that there is no significant difference between the models they connect.}
    \label{fig:diagram_cd}
\end{figure}

The diagram thus complements the fold-level results by showing that the observed performance hierarchy is not merely numerical but statistically supported, while also highlighting clusters of methods whose differences are not statistically meaningful. 
In the context of review-aware recommender systems, textual signals have repeatedly been shown to improve rating prediction under specific settings. In recommendation pipelines, the goal is not just to minimize pointwise error but also to produce high-quality rankings. Despite significant advances in \gls{nlp}, leveraging reviews as model inputs remains non-trivial. The mechanisms by which text improves recommendations are often indirect, dataset-dependent, and difficult to diagnose.

\begin{figure}
    \centering
    \includegraphics[width=1\linewidth]{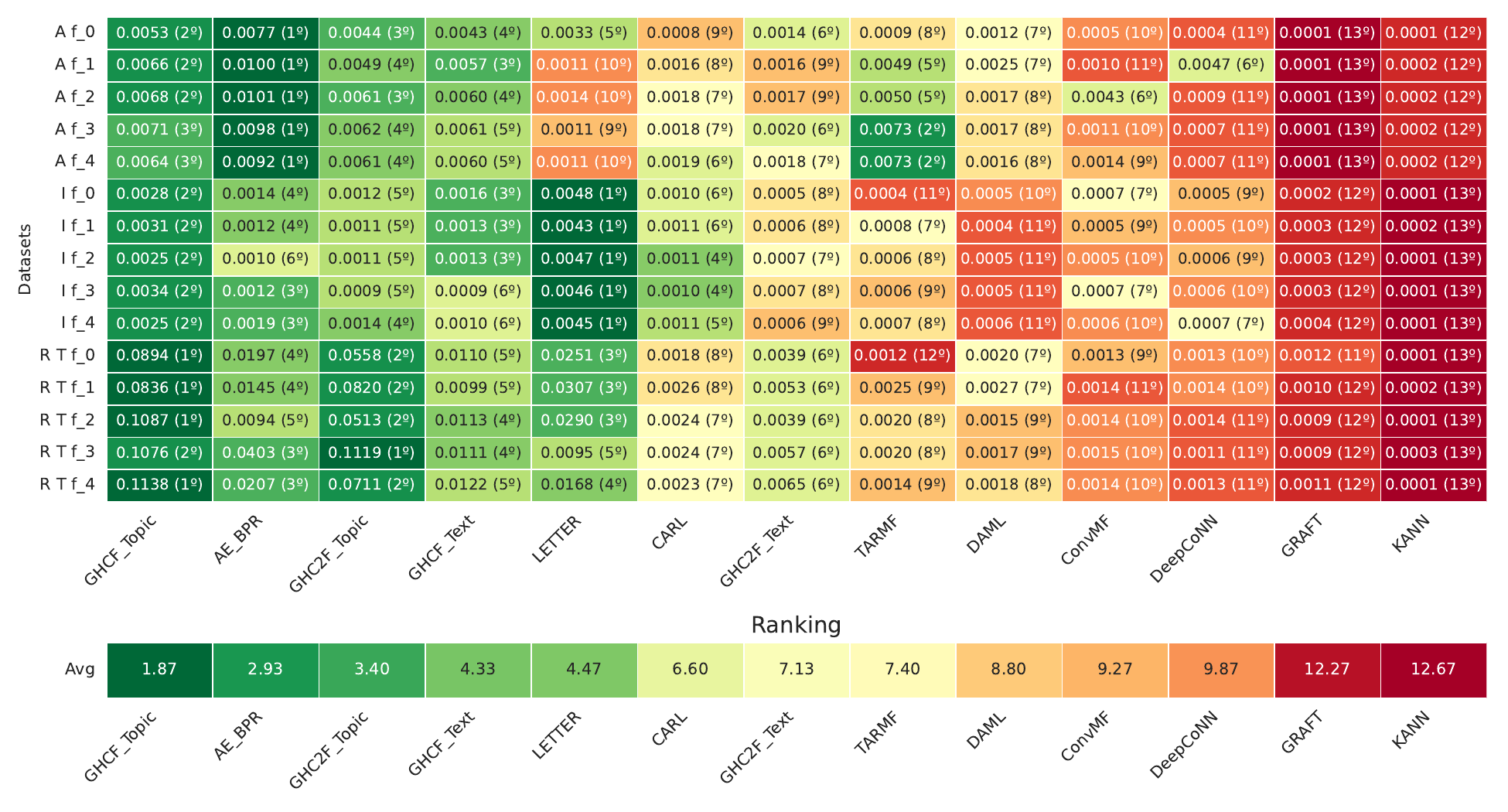}
    \caption{Comparison of the average ratings of nine recommendation algorithms across three datasets (A = Amazon, I = IMDb, and R T = Rotten Tomatoes) in different folds for the hypervolume metric. The last row presents the Global Average Rating of the algorithms, offering a fair non-parametric performance hierarchy that avoids the pitfalls of averaging incommensurable raw scores.}
    \label{fig:heatmap_geral}
\end{figure}

From a modeling perspective, there is a clear trade-off between expressiveness, computational cost, and interpretability. Architectures that rely on \gls{cnn}-based encoders to identify salient text spans can be computationally expensive, particularly when processing long reviews at scale. In contrast, approaches that explicitly extract aspects, sentiments, or topics tend to be more human-interpretable, since these intermediate representations can be inspected directly. Yet once such signals are injected into deep learning models, the causal pathway from text features to improvements in predictions becomes unclear. Consequently, it is difficult to determine whether gains reflect meaningful semantic supervision or simply additional capacity and regularization effects.

\subsection{Comparative Discussion of Proposed Models}

Our primary objective is to evaluate how the backbone autoencoder, optimized via the \gls{bpr} loss, interacts with two key components: the layer-wise gating mechanism and the contrastive learning objective. To ensure a comprehensive evaluation, these variations are benchmarked using both latent topic distributions and dense textual embeddings, the latter generated with the pre-trained \texttt{all-mpnet-base-v2} transformer model. This analysis addresses specific gaps in the current literature. Unlike existing \gls{sota} models, which often treat review integration as an inseparable feature, our study includes an ablation version that disables the review module entirely. This allows for a precise measurement of the gain provided by textual side information. 

Furthermore, many prominent studies lack rigorous statistical validation, often failing to specify whether reported metrics are based on a single execution or on robust $k$-fold cross-validation. In the absence of significance testing, it becomes impossible to distinguish true architectural superiority from stochastic variance. Consequently, comparing gains without a rigorous experimental framework constitutes a naive analysis, as such outcomes cannot be reliably disentangled from the probabilistic noise inherent in these models.

\begin{figure}[ht]
    \centering
    \includegraphics[width=1\linewidth]{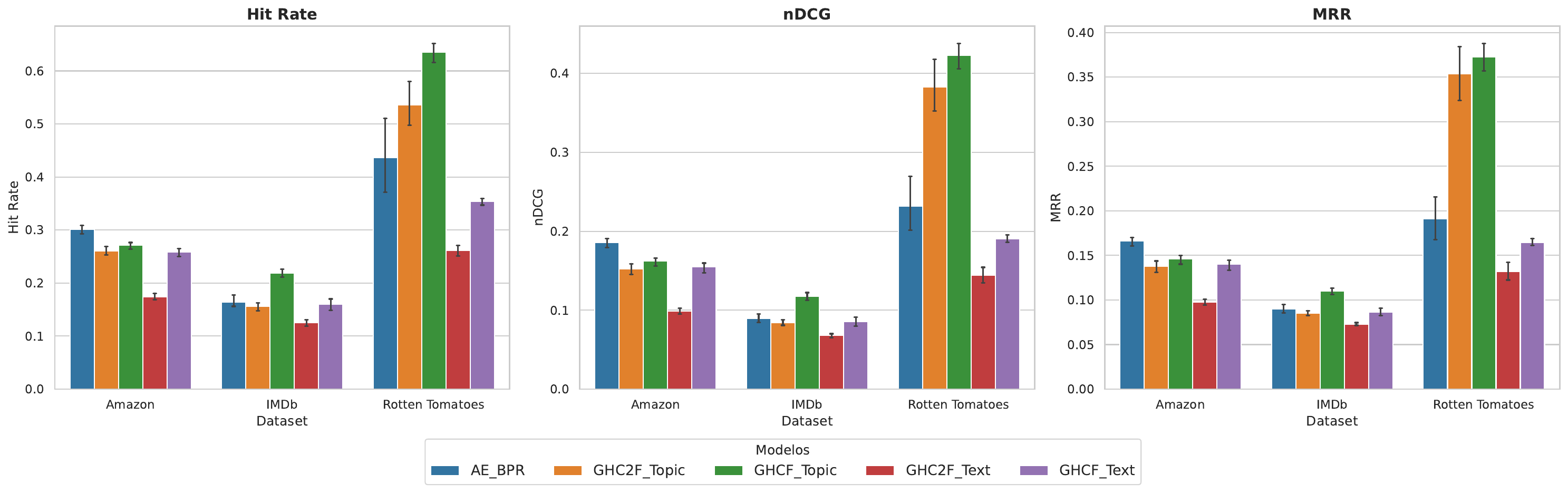}
    \caption{Detailed and consolidated ranking analysis across all datasets and folds. The top heatmap displays the performance rank of each model for each test block, while the bottom row represents the Global Average Ranking. GHCF\_Topic is the most stable performer, followed by the baseline AE\_BPR, whereas the embedding-based variants are more sensitive to specific dataset characteristics.}
    \label{fig:bar_metrics}
\end{figure}

Complementing the statistical discussion from the previous section, Figure~\ref{fig:bar_metrics} illustrates the performance consistency across the global average of all datasets. The superiority of the GHCF\_Topic model (represented in green) is visually evident. This indicates a consistent dominance across all three evaluation metrics: \gls{hr}, \gls{ndcg}, and \gls{mrr}. Effectively encompasses the performance frontiers of the other proposed variations and the AE\_BPR baseline. Additionally, Topic-based variations (\gls{ghcf}\_Topic, \gls{ghc2f}\_Topic) significantly outperform the text-based (Embedding) versions. For example, on the IMDb dataset, the \gls{ghcf}\_Topic achieves a global mean \gls{hr} of 0.22, while the embedding version (\gls{ghcf}\_Text) stays at 0.16. For these specific datasets (Amazon, IMDb, Rotten Tomatoes), extracted topics provide a more structured, noise-free representation of item features than dense embeddings do. Dense embeddings may capture too much semantic nuance that isn't directly relevant to the user’s preference signal, whereas topics align more closely with ``genres'', ``categories'', and aspects of the movies' characteristics that drive recommendations.

The transition from the baseline AE\_BPR to the gated proposal \gls{ghcf} yields the largest performance gain. On the Rotten Tomatoes dataset,  \gls{ghcf}\_Topic improves the Hit Rate by ~45\% over the AE\_BPR baseline (0.63 vs 0.43). This confirms that simply having an autoencoder is not enough. The Gating Mechanism acts as a ``filter'' that effectively weights the textual information. In other words, unlike traditional hybrid models that merely augment the feature vector, our approach ``decides'' which specific textual components are relevant for refining the latent user-item representations, effectively suppressing noise while amplifying high-utility signals.

While \gls{ghcf} achieves high raw performance across standard metrics, the \gls{ghc2f} (Contrastive) variant is specifically engineered to mitigate the challenges posed by data sparsity. Even when the mean performance across certain folds is marginally lower, the introduction of contrastive loss serves as a regularizer for the latent space. By maximizing the mutual information between the interaction and textual views, \gls{ghc2f} enforces a more structured and semantically aligned latent manifold. Although this study does not explicitly benchmark cold-start scenarios, the architectural design of \gls{ghc2f} is theoretically optimized for items with minimal interaction history, where textual signals must compensate for missing rating data. Consequently, while \gls{ghcf} excels in general recommendation ranking, \gls{ghc2f} provides the structural alignment essential to maintaining robustness in sparse-interaction environments.

\subsection{Limitations and improvements}

Despite its effectiveness, the proposed model has limitations. First, although it employs a gated fusion mechanism, alternative strategies, such as multi-head or hierarchical attention, could enable more expressive interactions between collaborative and textual representations, albeit at increased computational cost. Additionally, sentiment polarity and intensity are not explicitly modeled, limiting the use of opinion strength in ranking decisions. Second, topic representations are generated externally and are not optimized jointly with the recommender, making performance dependent on topic quality. Finally, the evaluation focuses on ranking accuracy, leaving explainability, fairness, and robustness to extreme sparsity underexplored. 

\section{Conclusion and Future Works}
\label{sec:conclusion}

This study investigated the impact of gated hybrid architectures and contrastive alignment on review-based recommendation systems. By benchmarking variations of the AE-BPR backbone against gated (GHCF) and contrastive (GHC2F) extensions, we addressed the critical question of whether structured alignment of rating-based collaborative representations and review-derived topic profiles improves ranking performance. 

Our findings, grounded in a rigorous experimental framework using $k$-fold cross-validation, reveal a clear performance hierarchy. Specifically, the GHCF\_Topic variant demonstrated superior consistency and robustness, as evidenced by its dominant area in the multi-objective radar analysis and its top ranking in the Friedman statistical test. These results indicate that adaptive layer-wise gating, particularly when applied to latent topic distributions, more effectively integrates behavioral and textual signals than simple feature fusion or dense embedding approaches. 

Furthermore, this work critiques the lack of statistical rigor in current state-of-the-art methodologies. By employing post hoc Nemenyi tests to validate performance gains, we move beyond the ``naive analysis'' of marginal percentage improvements, demonstrating that the observed improvements are statistically significant ($p < 0.05$) and not merely the result of stochastic noise. Despite these contributions, the approach exhibits sensitivity to review density and depends on the quality of the initial topic extraction. While the researcher-led qualitative audit ensured semantic relevance, the gating dynamics remain partially opaque. 

Future research should explore more sophisticated textual encoders, such as transformer-based architectures fine-tuned for recommendation, and investigate explicit constraints to improve the interpretability of the gating mechanism. Additionally, developing sparsity-aware strategies will be essential to further enhance the robustness and generalization of hybrid recommendation systems in cold-start scenarios.

\backmatter





\bmhead{Acknowledgements}


\bibliography{sn-bibliography}

\end{document}

%% file: acronym.tex
\newacronym{rbrs}{RBRS}{Review-based Recommender Systems}
\newacronym{rs}{RS}{Recommender System}
\newacronym{cf}{CF}{Collaborative Filtering}
\newacronym{mf}{MF}{Matrix Factorization}
\newacronym{nlp}{NLP}{Natural Language Processing}    
\newacronym{tfidf}{TF-IDF}{Term Frequency-Inverse Document Frequency}
\newacronym{bow}{BoW}{Bag-of-Words}
\newacronym{ai}{AI}{Artificial Intelligence}
\newacronym{ml}{ML}{Machine Learning}
\newacronym{mlp}{MLP}{Multi Layer Perceptron}
\newacronym{gat}{GAT}{Graph Attention Network}
\newacronym{nn}{NN}{Neural Network}
\newacronym{gcn}{GCN} {Graph Convolutional Network}
\newacronym{gnn}{GNN} {Graph Neural Network}
\newacronym{cnn}{CNN}{Convolutional Neural Network}
\newacronym{rnn}{RNN}{Recurrent Neural Network}
\newacronym{lda}{LDA}{Latent Dirichlet Allocation}
\newacronym{nmf}{NMF}{Non-Negative Matrix Factorization}
\newacronym{umap}{UMAP}{Uniform Manifold Approximation and Projection}
\newacronym{c-tf-idf}{c-TF-IDF}{class Term Frequency-Inverse Document Frequency}
\newacronym{hdbscan}{HDBSCAN}{Hierarchical Density-Based Spatial Clustering of Applications with Noise}
\newacronym{gte}{GTE}{thenlper/gte-small}
\newacronym{ncf}{NCF}{Neural Collaborative Filtering}
\newacronym{ngcf}{NGCF}{Neural Graph Collaborative Filtering}
\newacronym{svd}{SVD}{Singular Value Decomposition}
\newacronym{absa}{ABSA}{Aspect-based sentiment Analysis}
\newacronym{bert}{BERT}{Bidirectional Encoder Representations from Transformers}
\newacronym{roberta}{RoBERTa}{Robustly Optimized BERT Pretraining Approach}
\newacronym{rmse}{RMSE}{Root Mean Squared Error}
\newacronym{mae}{MAE}{Mean Absolute Error}
\newacronym{mse}{MSE} {Mean Squared Error}
\newacronym{map}{MAP}{Mean Average Precision}
\newacronym{mrr}{MRR}{Mean Reciprocal Rank}
\newacronym{gru}{GRU} {Gated Recurrent Unit}
\newacronym{lstm}{LSTM}{Long Short-Term Memory}
\newacronym{ndcg}{nDCG}{Normalized Discounted Cumulative Gain}
\newacronym{hr}{HR}{Hit Rate}
\newacronym{loo}{LOO}{Leave-One-Out}
\newacronym{cd}{CD}{Critical Difference}
\newacronym{hv}{HV}{Hypervolume}
\newacronym{bpr}{BPR}{Bayesian Personalized Ranking}
\newacronym{llm}{LLM}{Large Language Models}
\newacronym{ghcf}{GHCF}{Gated Hybrid Collaborative Filtering}
\newacronym{ghc2f}{GHC2F}{Gated Hybrid Contrastive Collaborative Filtering}
\newacronym{mmse}{MMSE}{Masked Mean Squared Error}
\newacronym{sota}{SOTA}{State of the Art}

%% file: sn-bibliography.bib
@ARTICLE{Tao_Chen_2022,
AUTHOR={Chen, Tao  and Samaranayake, Premaratne  and Cen, XiongYing  and Qi, Meng  and Lan, Yi-Chen },         
TITLE={The Impact of Online Reviews on Consumers’ Purchasing Decisions: Evidence From an Eye-Tracking Study},   
JOURNAL={Frontiers in Psychology},    
VOLUME={Volume 13 - 2022},
YEAR={2022},
URL={https://www.frontiersin.org/journals/psychology/articles/10.3389/fpsyg.2022.865702},
DOI={10.3389/fpsyg.2022.865702},
ISSN={1664-1078},
}

@Article{Zheng_Li_2023,
author={Li, Zheng
and Jin, Di
and Yuan, Ke},
title={Attentional factorization machine with review-based user--item interaction for recommendation},
journal={Scientific Reports},
year={2023},
month={Aug},
day={18},
volume={13},
number={1},
pages={13454},
issn={2045-2322},
doi={10.1038/s41598-023-40633-4},
url={https://doi.org/10.1038/s41598-023-40633-4}
}

@article{Qiang_Wang_2022,
title = {Effect of online review sentiment on product sales: The moderating role of review credibility perception},
journal = {Computers in Human Behavior},
volume = {133},
pages = {107272},
year = {2022},
issn = {0747-5632},
doi = {https://doi.org/10.1016/j.chb.2022.107272},
url = {https://www.sciencedirect.com/science/article/pii/S0747563222000942},
author = {Qiang Wang and Wen Zhang and Jian Li and Feng Mai and Zhenzhong Ma}
}

@misc{Xi_Wang_2021,
      title={Leveraging Review Properties for Effective Recommendation}, 
      author={Xi Wang and Iadh Ounis and Craig Macdonald},
      year={2021},
      eprint={2102.03089},
      archivePrefix={arXiv},
      primaryClass={cs.IR},
      url={https://arxiv.org/abs/2102.03089}, 
}

@ARTICLE{Adomavicius_2005,
  author={Adomavicius, G. and Tuzhilin, A.},
  journal={IEEE Transactions on Knowledge and Data Engineering}, 
  title={Toward the next generation of recommender systems: a survey of the state-of-the-art and possible extensions}, 
  year={2005},
  volume={17},
  number={6},
  pages={734-749},
  doi={10.1109/TKDE.2005.99}
}

@article{Khan_2023,
author = {Khan, N. Zafar Ali and Mahalakshmi, R.},
title = {A novel user review-based contextual recommender system},
journal = {International Journal of Modeling, Simulation, and Scientific Computing},
volume = {14},
number = {01},
pages = {2341002},
year = {2023},
doi = {10.1142/S1793962323410027},
URL = {https://doi.org/10.1142/S1793962323410027},
eprint = {https://doi.org/10.1142/S1793962323410027}
}

@Article{Li_Chen_2015,
author={Chen, Li
and Chen, Guanliang
and Wang, Feng},
title={Recommender systems based on user reviews: the state of the art},
journal={User Modeling and User-Adapted Interaction},
year={2015},
month={Jun},
day={01},
volume={25},
number={2},
pages={99-154},
issn={1573-1391},
doi={10.1007/s11257-015-9155-5},
url={https://doi.org/10.1007/s11257-015-9155-5}
}

@article{Gheewala_2024,
title = {Exploiting deep transformer models in textual review based recommender systems},
journal = {Expert Systems with Applications},
volume = {235},
pages = {121120},
year = {2024},
issn = {0957-4174},
doi = {https://doi.org/10.1016/j.eswa.2023.121120},
url = {https://www.sciencedirect.com/science/article/pii/S0957417423016226},
author = {Shivangi Gheewala and Shuxiang Xu and Soonja Yeom and Sumbal Maqsood}
}

@Article{Zhuang_2021,
AUTHOR = {Zhuang, Yuanyuan and Kim, Jaekyeong},
TITLE = {A BERT-Based Multi-Criteria Recommender System for Hotel Promotion Management},
JOURNAL = {Sustainability},
VOLUME = {13},
YEAR = {2021},
NUMBER = {14},
ARTICLE-NUMBER = {8039},
ISSN = {2071-1050},
DOI = {10.3390/su13148039}
}

@Article{Raza_2022,
author={Raza, Shaina
and Ding, Chen},
title={News recommender system: a review of recent progress, challenges, and opportunities},
journal={Artificial Intelligence Review},
year={2022},
month={Jan},
day={01},
volume={55},
number={1},
pages={749-800},
issn={1573-7462},
doi={10.1007/s10462-021-10043-x},
url={https://doi.org/10.1007/s10462-021-10043-x}
}

@inproceedings{Chong_Chen_2018,
author = {Chen, Chong and Zhang, Min and Liu, Yiqun and Ma, Shaoping},
title = {Neural Attentional Rating Regression with Review-level Explanations},
year = {2018},
isbn = {9781450356398},
publisher = {International World Wide Web Conferences Steering Committee},
address = {Republic and Canton of Geneva, CHE},
doi = {10.1145/3178876.3186070},
booktitle = {Proceedings of the 2018 World Wide Web Conference},
pages = {1583–1592},
numpages = {10},
location = {Lyon, France},
series = {WWW '18}
}

@inproceedings{Zheng_2017,
author = {Zheng, Lei and Noroozi, Vahid and Yu, Philip S.},
title = {Joint Deep Modeling of Users and Items Using Reviews for Recommendation},
year = {2017},
isbn = {9781450346757},
publisher = {Association for Computing Machinery},
address = {New York, NY, USA},
doi = {10.1145/3018661.3018665},
booktitle = {Proceedings of the Tenth ACM International Conference on Web Search and Data Mining},
pages = {425–434},
numpages = {10},
location = {Cambridge, United Kingdom},
series = {WSDM '17}
}

@inproceedings{Zhiyong_Cheng_2028,
author = {Cheng, Zhiyong and Ding, Ying and He, Xiangnan and Zhu, Lei and Song, Xuemeng and Kankanhalli, Mohan},
title = {A3NCF: an adaptive aspect attention model for rating prediction},
year = {2018},
isbn = {9780999241127},
address = {Stockholm, Sweden},
publisher = {AAAI Press},
booktitle = {Proceedings of the 27th International Joint Conference on Artificial Intelligence},
doi = {https://dl.acm.org/doi/10.5555/3304222.3304290},
pages = {3748–3754},
numpages = {7},
location = {Stockholm, Sweden},
series = {IJCAI'18}
}

@inproceedings{Chin_2018,
author = {Chin, Jin Yao and Zhao, Kaiqi and Joty, Shafiq and Cong, Gao},
title = {ANR: Aspect-based Neural Recommender},
year = {2018},
isbn = {9781450360142},
publisher = {Association for Computing Machinery},
address = {New York, NY, USA},
doi = {10.1145/3269206.3271810},
booktitle = {Proceedings of the 27th ACM International Conference on Information and Knowledge Management},
pages = {147–156},
numpages = {10},
location = {Torino, Italy},
series = {CIKM '18}
}

@inproceedings{Musto_2017,
author = {Musto, Cataldo and de Gemmis, Marco and Semeraro, Giovanni and Lops, Pasquale},
title = {A Multi-criteria Recommender System Exploiting Aspect-based Sentiment Analysis of Users' Reviews},
year = {2017},
isbn = {9781450346528},
publisher = {Association for Computing Machinery},
address = {New York, NY, USA},
doi = {10.1145/3109859.3109905},
booktitle = {Proceedings of the Eleventh ACM Conference on Recommender Systems},
pages = {321–325},
numpages = {5},
location = {Como, Italy},
series = {RecSys '17}
}

@inproceedings{Zhongxia_Chen_2019,
author = {Chen, Zhongxia and Wang, Xiting and Xie, Xing and Wu, Tong and Bu, Guoqing and Wang, Yining and Chen, Enhong},
title = {Co-attentive multi-task learning for explainable recommendation},
address = {Macao, China},
year = {2019},
isbn = {9780999241141},
publisher = {AAAI Press},
booktitle = {Proceedings of the 28th International Joint Conference on Artificial Intelligence},
pages = {2137–2143},
numpages = {7},
location = {Macao, China},
series = {IJCAI'19}
}

@inproceedings{Liu_2019,
author = {Liu, Donghua and Li, Jing and Du, Bo and Chang, Jun and Gao, Rong},
title = {DAML: Dual Attention Mutual Learning between Ratings and Reviews for Item Recommendation},
year = {2019},
isbn = {9781450362016},
publisher = {Association for Computing Machinery},
doi = {10.1145/3292500.3330906},
booktitle = {Proceedings of the 25th ACM SIGKDD International Conference on Knowledge Discovery \& Data Mining},
pages = {344–352},
numpages = {9},
location = {Anchorage, AK, USA},
address = {Anchorage, AK, USA},
series = {KDD '19}
}

@inproceedings{Shuai_2022,
author = {Shuai, Jie and Zhang, Kun and Wu, Le and Sun, Peijie and Hong, Richang and Wang, Meng and Li, Yong},
title = {A Review-aware Graph Contrastive Learning Framework for Recommendation},
year = {2022},
isbn = {9781450387323},
publisher = {Association for Computing Machinery},
address = {New York, NY, USA},
doi = {10.1145/3477495.3531927},
booktitle = {Proceedings of the 45th International ACM SIGIR Conference on Research and Development in Information Retrieval},
pages = {1283–1293},
numpages = {11},
location = {Madrid, Spain},
series = {SIGIR '22}
}

@article{Shang_2024,
  author       = {Fu Shang and Jiatu Shi and Yadong Shi and Shuwen Zhou},
  title        = {Enhancing E-Commerce Recommendation Systems with Deep Learning-based Sentiment Analysis of User Reviews},
  journal      = {International Journal of Engineering and Management Research},
  year         = 2024,
  volume       = 14,
  number       = 4,
  month        = aug,
  doi          = {10.5281/zenodo.13221409},
}

@article{Bittencourt_2025,
author = {Bittencourt, Guilherme and Vasconcelos, Naan and Andrade, Yan and Silva, N\'{\i}collas and Cunha, Washington and Colombo Dias, Diego Roberto and Gon\c{c}alves, Marcos Andr\'{e} and Rocha, Leonardo},
title = {Review-Aware Recommender Systems (RARSs): Recent Advances, Experimental Comparative Analysis, Discussions, and New Directions},
year = {2025},
issue_date = {January 2026},
publisher = {Association for Computing Machinery},
address = {New York, NY, USA},
volume = {58},
number = {1},
issn = {0360-0300},
url = {https://doi.org/10.1145/3744661},
doi = {10.1145/3744661},
journal = {ACM Comput. Surv.},
month = sep,
articleno = {21},
numpages = {49}
}

@article{Libing_Wu_2019,
author = {Wu, Libing and Quan, Cong and Li, Chenliang and Wang, Qian and Zheng, Bolong and Luo, Xiangyang},
title = {A Context-Aware User-Item Representation Learning for Item Recommendation},
year = {2019},
issue_date = {April 2019},
publisher = {Association for Computing Machinery},
address = {New York, NY, USA},
volume = {37},
number = {2},
issn = {1046-8188},
url = {https://doi.org/10.1145/3298988},
doi = {10.1145/3298988},
journal = {ACM Trans. Inf. Syst.},
month = jan,
articleno = {22},
numpages = {29}
}

@inproceedings{Kim_2016,
author = {Kim, Donghyun and Park, Chanyoung and Oh, Jinoh and Lee, Sungyoung and Yu, Hwanjo},
title = {Convolutional Matrix Factorization for Document Context-Aware Recommendation},
year = {2016},
isbn = {9781450340359},
publisher = {Association for Computing Machinery},
address = {New York, NY, USA},
doi = {10.1145/2959100.2959165},
booktitle = {Proceedings of the 10th ACM Conference on Recommender Systems},
pages = {233–240},
numpages = {8},
location = {Boston, Massachusetts, USA},
series = {RecSys '16}
}

@inproceedings{Eduardo_2025,
 author = {Eduardo Silva and Joel Pires and Denis Dantas and Mayki Oliveira and Frederico Durão},
 title = { A Gated Review Attention Framework for Topics in Graph-Based Recommenders},
 booktitle = {Proceedings of the 31st Brazilian Symposium on Multimedia and the Web},
 location = {Rio de Janeiro/RJ},
 year = {2025},
 issn = {0000-0000},
 pages = {19--27},
 publisher = {SBC},
 address = {Porto Alegre, RS, Brasil},
 doi = {10.5753/webmedia.2025.15515}
}

@Article{Liu_2023,
author={Liu, Yun
and Miyazaki, Jun},
title={Knowledge-aware attentional neural network for review-based movie recommendation with explanations},
journal={Neural Computing and Applications},
year={2023},
month={Jan},
day={01},
volume={35},
number={3},
pages={2717-2735},
issn={1433-3058},
doi={10.1007/s00521-022-07689-1},
url={https://doi.org/10.1007/s00521-022-07689-1}
}

@inproceedings{Jiwon_Son_2025,
author = {Son, Jiwon and Kim, Hyunjoon and Kim, Sang-Wook},
title = {Rating-Aware Homogeneous Review Graphs and User Likes/Dislikes Differentiation for Effective Recommendations},
year = {2025},
isbn = {9798400715921},
publisher = {Association for Computing Machinery},
address = {New York, NY, USA},
doi = {10.1145/3726302.3730069},
booktitle = {Proceedings of the 48th International ACM SIGIR Conference on Research and Development in Information Retrieval},
pages = {2070–2080},
numpages = {11},
location = {Padua, Italy},
series = {SIGIR '25}
}

@inproceedings{Yichao_Lu_2018,
author = {Lu, Yichao and Dong, Ruihai and Smyth, Barry},
title = {Coevolutionary Recommendation Model: Mutual Learning between Ratings and Reviews},
year = {2018},
isbn = {9781450356398},
publisher = {International World Wide Web Conferences Steering Committee},
address = {Republic and Canton of Geneva, CHE},
url = {https://doi.org/10.1145/3178876.3186158},
doi = {10.1145/3178876.3186158},
booktitle = {Proceedings of the 2018 World Wide Web Conference},
pages = {773–782},
numpages = {10},
location = {Lyon, France},
series = {WWW '18}
}

@article{Demsar_2006,
author = {Dem\v{s}ar, Janez},
title = {Statistical Comparisons of Classifiers over Multiple Data Sets},
year = {2006},
issue_date = {12/1/2006},
publisher = {JMLR.org},
volume = {7},
issn = {1532-4435},
journal = {J. Mach. Learn. Res.},
month = dec,
pages = {1–30},
numpages = {30}
}

@article{li2023towards,
  title={Towards general text embeddings with multi-stage contrastive learning},
  author={Li, Zehan and Zhang, Xin and Zhang, Yanzhao and Long, Dingkun and Xie, Pengjun and Zhang, Meishan},
  journal={arXiv preprint arXiv:2308.03281},
  year={2023}
}

@article{Tan2025,
  title={Do Reviews Matter for Recommendations in the Era of Large Language Models?},
  author={Chee Heng Tan and Huiying Zheng and Jing Wang and Zhuoyi Lin and Shaodi Feng and Huijing Zhan and Xiaoli Li and J. Senthilnath},
  journal={ArXiv},
  year={2025},
  volume={abs/2512.12978},
  url={https://api.semanticscholar.org/CorpusID:283895950}
}

@misc{grootendorst2022,
      title={BERTopic: Neural topic modeling with a class-based TF-IDF procedure}, 
      author={Maarten Grootendorst},
      year={2022},
      eprint={2203.05794},
      archivePrefix={arXiv},
      primaryClass={cs.CL},
      url={https://arxiv.org/abs/2203.05794}, 
}

@inproceedings{coherence,
author = {R\"{o}der, Michael and Both, Andreas and Hinneburg, Alexander},
title = {Exploring the Space of Topic Coherence Measures},
year = {2015},
isbn = {9781450333177},
publisher = {Association for Computing Machinery},
address = {New York, NY, USA},
url = {https://doi.org/10.1145/2684822.2685324},
doi = {10.1145/2684822.2685324},
booktitle = {Proceedings of the Eighth ACM International Conference on Web Search and Data Mining},
pages = {399–408},
numpages = {10},
location = {Shanghai, China},
series = {WSDM '15}
}

@misc{infoce_2019,
      title={Representation Learning with Contrastive Predictive Coding}, 
      author={Aaron van den Oord and Yazhe Li and Oriol Vinyals},
      year={2019},
      eprint={1807.03748},
      archivePrefix={arXiv},
      primaryClass={cs.LG},
      url={https://arxiv.org/abs/1807.03748}, 
}

@article{irev_2025,
author = {Bittencourt, Guilherme and Vasconcelos, Naan and Andrade, Yan and Silva, N\'{\i}collas and Cunha, Washington and Colombo Dias, Diego Roberto and Gon\c{c}alves, Marcos Andr\'{e} and Rocha, Leonardo},
title = {Review-Aware Recommender Systems (RARSs): Recent Advances, Experimental Comparative Analysis, Discussions, and New Directions},
year = {2025},
issue_date = {January 2026},
publisher = {Association for Computing Machinery},
address = {New York, NY, USA},
volume = {58},
number = {1},
issn = {0360-0300},
url = {https://doi.org/10.1145/3744661},
doi = {10.1145/3744661},
journal = {ACM Comput. Surv.},
month = sep,
articleno = {21},
numpages = {49}
}
